# Focused Surface Acoustic Wave induced nano-oscillator based reservoir computing

Md. Fahim F. Chowdhury[1], Walid Al Misba[1], Md Mahadi Rajib[1], Alexander J. Edwards[2], Dhritiman Bhattacharya[3], Joseph S. Friedman[2], Jayasimha Atulasimha[1*]

[1]Department of Mechanical and Nuclear Engineering, Virginia Commonwealth University, Richmond, VA
[1]{chowdhurymf, misbawa, rajibmm, *jatulasimha}@vcu.edu
[2]Department of Electrical and Computer Engineering, The University of Texas at Dallas, Richardson, TX
[2]{alexander.edwards, joseph.friedman}@utdallas.edu
[3]Department of Physics, Georgetown University, Washington, DC
[3]dhritiman.bhattacharya@georgetown.edu

We demonstrate using micromagnetic simulations that a nanomagnet array excited by Surface Acoustic Waves (SAWs) can work as a reservoir. An input nanomagnet is excited with focused SAW and coupled to several nanomagnets, seven of which serve as output nanomagnets. To evaluate memory effect and computing capability, we study the Short-Term Memory (STM) and Parity Check (PC) capacities respectively. The SAW has a carrier frequency of 4 GHz whose amplitude is modulated to provide different inputs of sine and square waves of 100 MHz frequency. The responses of the selected output nanomagnets are processed by reading the envelope of their magnetization state, which is used to train the output weights using regression method (e.g. Moore-Penrose pseudoinverse operation). For classification, a random sequence of 100 square and sine wave samples are used, of which 80 % are used for training, and the rest of the samples are used for testing. We achieve 100 % training accuracy and 100 % testing accuracy. Furthermore, the average STM and PC are calculated to be ~4.69 bits and ~5.39 bits respectively, which is indicative of the proposed acoustically driven nanomagnet oscillator array being well suited for physical reservoir computing applications. The energy dissipation is two-orders lower than a CMOS-based echo-state network. Finally, the ability to use high frequency (4 GHz) SAW makes the device scalable to small dimensions, while the ability to modulate the envelope at a lower frequency (100 MHz) adds flexibility to encode different signals beyond the sine and square waves demonstrated here.

A Recurrent Neural Network (RNN) is a machine learning algorithm, which uses its internal memory to remember previous inputs and hence process time-series data e.g., speech, audio, text, weather, etc. Reservoir Computing (RC) is derived from the RNN theory and is a computational framework where a fixed, non-linear reservoir maps the inputs into higher-dimensional space and the readout is trained with linear regression and classification[1]. A RC network consists of inputs, reservoirs, and outputs as shown in Fig. 1(a). In a RC network, only the output weights are trained with a fast and simple linear regression method, which enables the implementation of efficient training. Such physical reservoir implementations are suitable for edge devices that need to learn in real-time with limited hardware, computational resources, and energy. An ideal physical reservoir should have short-term memory effect and non-linear dynamics as well as be amenable to manufacturing with minimal circuitry. Various Physical RC (PRC) systems are proposed by researchers such as spintronic PRC[2-12], electronic PRC[2], photonic PRC[13-14], etc. Each of these physical reservoirs has respective advantages and disadvantages.

Spintronic nanomagnetic devices are particularly well suited for physical reservoir computing due to their inherent interactive non-linear dynamics, recurrence characteristics, enduring lifetime, CMOS-compatibility, and low energy consumption[2-3]. Spintronic PRC has been simulated or experimentally implemented using dipole-coupled nanomagnets[15-16], spin-torque nano-oscillators (STNOs)[10,17-18], spin-wave systems[19-21], and different skyrmion fabrics[4-6,22]. Simple pattern recognition task can be performed with a skyrmion fabric reservoir, which utilizes the recursive response of magnetization dynamics[5]. Complex tasks such as image classification can also be performed by a single magnetic skyrmion memristor (MSM) with current pulse stimulation[4,6]. Several studies have proposed domain wall (DW) based neurons and synapses for integrated hybrid CMOS and spintronic computing[7-9]. Apart from skyrmion textures and domain walls, vortex-type spin torque oscillator[10], magnetic-dipole interactions[15] can be used as a resource for nonlinear dynamics of a spintronic reservoir. Higher computational capabilities can be achieved using forced synchronization[10], by increasing the number of STNOs, or at the boundary between synchronized and disordered states[23].

Recently, strain-mediated nanomagnet devices[28-29] were demonstrated for memory applications through resonant surface acoustic wave (r-SAW) assisted spin-transfer-torque[24-25]. Unlike memory application, reservoir computing does not require the nanomagnets to switch to an orthogonal state or undergo a complete reversal. Hence, the energy barrier ($E_b = K_u V \sim 1 eV$) constraint, associated with volume (V), and perpendicular anisotropy constant ($K_u$) is not critical to its working. The SAW induced stress at a suitable frequency can induce ferromagnetic resonance, which leads to large amplitude precession while being energy efficient. These advantages motivated us to propose SAW induced magnetization dynamics as an input to nanomagnetic reservoirs. SAWs are generated by an inter-digitated transducer (IDT) patterned on a piezoelectric substrate, which produces Raleigh (transverse) waves. Piezoelectric materials such as Lithium Niobate ($LiNbO_3$), can be used to generate such SAW waves that induce magnetization dynamics in magnetostrictive nanomagnets fabricated on these substrates.

In this work, we demonstrate via micromagnetic simulation that a nanomagnet array, shown in Fig. 1, excited by surface acoustic wave (SAW) can be used as a reservoir to classify sine and square waves with high accuracy. We also evaluate two figures of merit tasks of RC named short-term memory (STM) capacity and parity check (PC) capacity. The STM and PC capacity tasks characterize the memory effect (influence of past states) and computing capability (non-linearity) of the system, respectively[26]. The amplitude of the SAW applied to the input nanomagnet is modulated in such a way that its envelope forms random sequence of sine and square waves of 100 MHz frequency. The non-linear responses of the output nanomagnets due to such an input are processed by reading the reservoir state in certain intervals and then trained to classify sine and square waves and calculate STM and PC capacity.

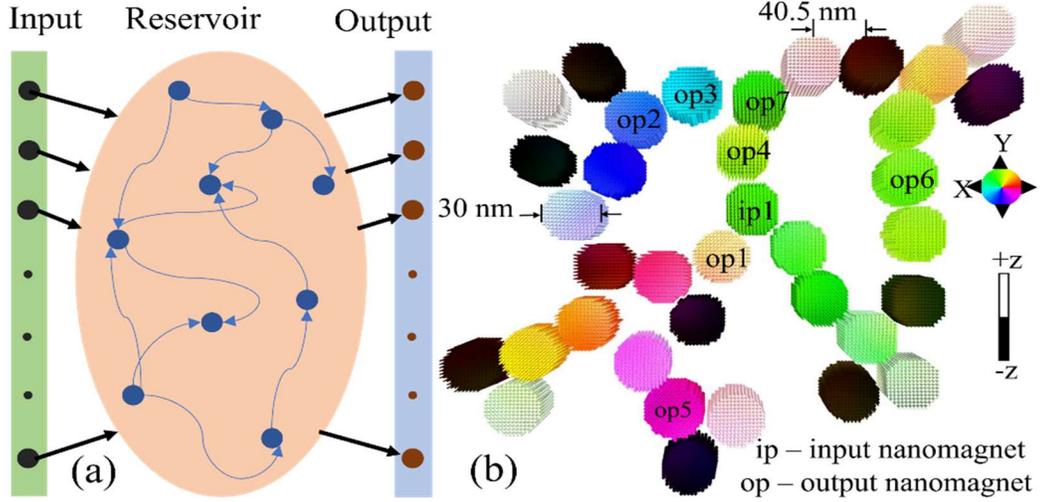

Fig. 1: (a) Concept of reservoir computing (b) A micro-magnetic snapshot of the input, the reservoir and output nanomagnets. The SAW is applied to the input nanomagnet (ip1) and the magnetizations of the output nanomagnets (op1 to op7) are read.

We obtain the free layer magnetization dynamics of the reservoir through micromagnetic simulation with MuMax3[27]. The magnetization direction of the reference ferromagnetic layer of an MTJ is fixed and the free layer magnetization is governed by the Landau-Lifshitz-Gilbert (LLG) equation as follows:

$$\frac{d\vec{m}}{dt} = -\frac{1}{(1+\alpha^2)}\gamma[\vec{m} \times \vec{H}_{effective}] - \frac{\alpha}{(1+\alpha^2)}\gamma[\vec{m} \times (\vec{m} \times \vec{H}_{effective})] \quad (1)$$

Here, $\vec{m}$ is the normalized magnetization defined as $\frac{\vec{M}}{M_s}$, $\vec{M}$ is the magnetization, $M_s$ is the saturation magnetization, $\alpha$ is the Gilbert damping coefficient, $\gamma$ is the gyromagnetic ratio. The effective magnetic field, $\vec{H}_{effective}$ is comprised of the fields due to the SAW induced stress anisotropy ($\vec{H}_{stress\ anis}$), demagnetizing field due to shape anisotropy ($\vec{H}_{demag}$), and the exchange field due to Heisenberg exchange coupling ($\vec{H}_{exchange}$) as described in the equation below.

$$\vec{H}_{effective} = \vec{H}_{stress\ anis} + \vec{H}_{demag} + \vec{H}_{exchange} \quad (2)$$

The effective field due to stress, $\vec{H}_{stress\ anis}$ (in the form of cyclic tension and compression)[24] due to the inverse magnetostriction effect[25] can be expressed as:

$$\vec{H}_{stress\ anis} = \frac{2K_{ut}}{\mu_0 M_s}(\vec{u}.\vec{m})\vec{u} \quad (3)$$

Here, $\mu_0$ is the magnetic permeability of free space, $\vec{u}$ is the applied stress direction. The stress anisotropy constant, $K_{ut}$ is $\frac{3}{2}\sigma\lambda_s$, where $\sigma$ is the induced stress by SAW and $\lambda_s$ is saturation magnetostriction. We consider a uniaxial stress induced by SAW in the $\vec{u}$ direction and neglect the in-plane component which experiences opposite stress orthogonal to $\vec{u}$ due to Poisson's effect. We note that the estimated stress amplitude is conservative due to this assumption, but the qualitative magnetization dynamics remain the same. Since the focused SAW is locally applied to the region of the input magnet only, the induced stress in the piezoelectric substrate in the reservoir or output region is comparatively negligible. So, the stress anisotropy field, $\vec{H}_{stress\ anis} = 0$ and the effective field on the nanomagnets of the reservoir or output nanomagnets is comprised of $\vec{H}_{demag}$ and $\vec{H}_{exchange}$.

$\vec{H}_{demag}$ is calculated by MuMax[27] at every point in each nanomagnet due to shape anisotropy of the nanomagnet itself and due to dipole coupling from other nanomagnets.

Finally, the Perpendicular Magnetic Anisotropy (PMA) is set to zero as PMA is negligible when the soft later thickness exceeds 1-2 nm and the thickness is 35 nm in our case.

The schematic diagram of the input, reservoir, and output nanomagnets for the RC simulation is shown in Fig. 1(b). The input nanomagnet is indicated by ip1 and of the rest of the nanomagnets in the reservoir, seven nanomagnets are selected as outputs and denoted by op1 to op7. A 4 GHz focused surface acoustic wave induced stress is applied to the input nanomagnet. We assume the SAW is applied using a focused interdigitated transducer (FIDT), which is patterned on top of a piezoelectric. The simulation dimension is 256 x 256 x 16 cells which covers all input, reservoir, and output nanomagnets, and each cell size is 2 nm x 2 nm x 2.1875 nm, which is much lower than the ferromagnetic exchange length, $\sqrt{2A_{ex}/\mu_0 M_s^2} = 6.32$ nm. The cylindrical nanomagnets are 30 nm in diameter and 35 nm in height. The piezoelectric substrate is assumed to be lithium niobate (LiNbO$_3$) and the simulation parameters[39-42] are summarized in Table I.

Table I. Simulation parameters for the physical reservoir for the soft ferromagnetic CoFe[39-42] layer.

| Parameter | Value |
| --- | --- |
| Gilbert damping constant, $\alpha$ | 0.05 |
| Saturation magnetization, $M_s$ | 0.72 x 10$^6$ A/m |
| Exchange stiffness, $A_{ex}$ | 13 x 10$^{-12}$ J/m |
| Free layer thickness, $t$ | 35 nm |
| Nanomagnet diameter, $D$ | 30 nm |

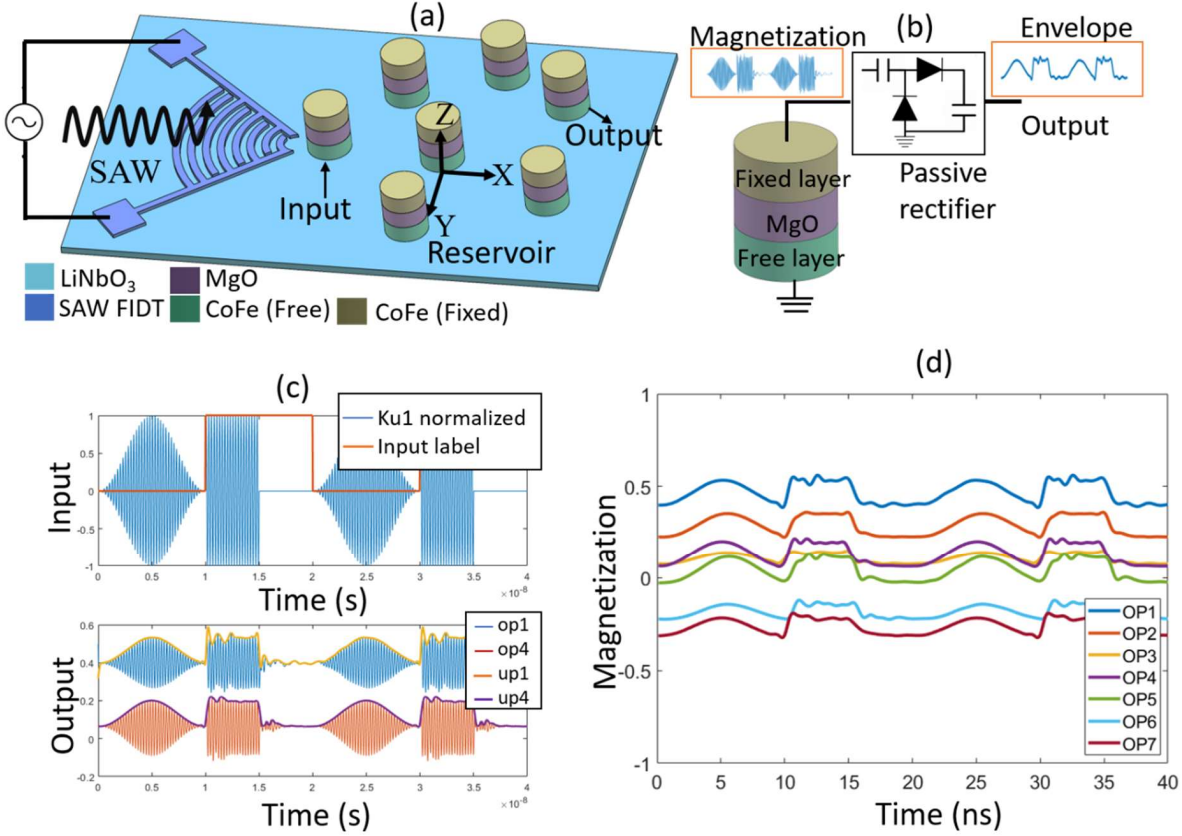

Fig. 2: (a) Simulation schematic showing application of SAW with Focused Inter-Digitated Transducer (FIDT) on the input nanomagnet as well as reservoir and output nanomagnets (b) Electrical readout of the magnetization of the output nanomagnet softlayer with an MTJ (c) Normalized stress anisotropy applied using SAW and labeling of sine and square waves as 1 and 0 (top), magnetization of output nanomagnets 1 and 4 in response to SAW and their corresponding envelopes (bottom) (d) The envelopes of the responses vs. time (ns) of several output nanomagnets.

Two fundamental properties required for reservoir computing are nonlinearity and memory[38]. Due to the nonlinearity and complex dynamics of the reservoir, the network response should be consistent/similar for similar inputs and distinguishable for different inputs[30]. For RC, we utilize the magnetization dynamics of the input and output nano-oscillators, which are governed by the LLG equation described earlier. Further, the input information is encoded in the envelope of a focused SAW of 4 GHz consisting of a random sequence of sine and square waves, applied to the input nanomagnet as shown in Fig. 2(a). During classification, the sine and square waves are labeled as 1 and 0, respectively. The details of the RC method are presented in the supplementary.

We evaluate the quantitative performance of the reservoir with STM task and PC task[30]. STM task characterizes the memory effect of the system by generating delayed inputs and testing if the internal dynamics of the reservoir is trained to adjust to this delay. The training and testing input data for STM is given below:

$$y_{STM}^{n,d} = i_s(n-d) \tag{4}$$

Here, $d$ = introduced delay. Since the STM task is not sufficient to prove reservoir property, the PC task is also evaluated as a benchmark task. The PC task characterizes the non-linearity of the system, which is indicative of the computing capability of the system and simplifies the training of the reservoir. The training and testing data for the PC task is prepared with modulo(2) operation to introduce non-linearity and is expressed as follows:

$$y_{PC}^{n,d} = [i_s(n-d) + i_s(n-d+1) + \cdots + i_s(n)] \, mod(2); \; d \neq 0 \quad (5)$$

Once the learned weights are obtained, the correlation coefficient between testing data, $y_{STM/PC}^{n,d}$ and output data $y_{out}$ are calculated. The total capacities for STM ($C_{STM}$) and PC ($C_{PC}$) tasks are calculated by integrating (summing in the discrete case) the correlation coefficients for delay up to $d_{max}$.

$$r_{STM/PC}(d) = \sqrt{\frac{covariance[y_{STM/PC}^{n,d}, y_{out}]}{variance[y_{STM/PC}^{n,d}] \, variance[y_{out}]}} \quad (6)$$

$$C_{STM/PC} = \sum_{d=0}^{d_{max}} [r_{STM/PC}(d)]^2 \quad (7)$$

Fig. 2(a) shows an example of the experimental setup of the proposed reservoir with the application of focused SAW. The focused SAW IDT and the reservoir are fabricated on a piezoelectric substrate, which is assumed to be lithium niobite (LiNbO$_3$). The input, reservoirs, and outputs are realized by magnetic tunnel junctions (MTJs), made of two CoFe layers (free layer and reference layer) separated by a tunnel barrier layer (MgO). The free layer magnetization responses are read from the output nanomagnets and preprocessed to obtain envelopes by spline interpolation[31] over local maxima separated by at least 3 samples. The upper envelopes of the output nanomagnets are shown in Fig. 2(d). Each sine or square signal is sampled into $N$ nodes separated by a sampling time $\tau$. The node density can be increased by introducing virtual nodes[32-33]. The signals are labeled as 0 and 1 in response to the sine and square waves, respectively. The weights are obtained by the linear regression method explained above.

To quantify the performance of the proposed reservoir, the sine and square wave classification is performed by the reservoir as a first task. Although simple, this classification task requires non-linearity and memory effects of the system to predict or classify these waves with high accuracy. The input is a random sequence of 100 sine and square waves with equal period of 10 ns. The first 80 signals are used to train, and the next 20 signals are used to test the reservoir for signal classification, STM task, and PC tasks. The reservoir is able to achieve 100 % training and 100 % testing accuracy with any of the output nanomagnets. The training and testing are performed for the different numbers of virtual nodes 5, 10, 20, 25, and 50, where 100 % recognition rate in both training and testing was achieved for all these numbers of nodes.

To further evaluate the performance of the reservoir, we studied two fundamental characteristics: fading memory and non-linearity. To evaluate the memory of the proposed reservoir we have calculated STM capacity and to evaluate the nonlinearity, we have performed the PC task and results are discussed next.

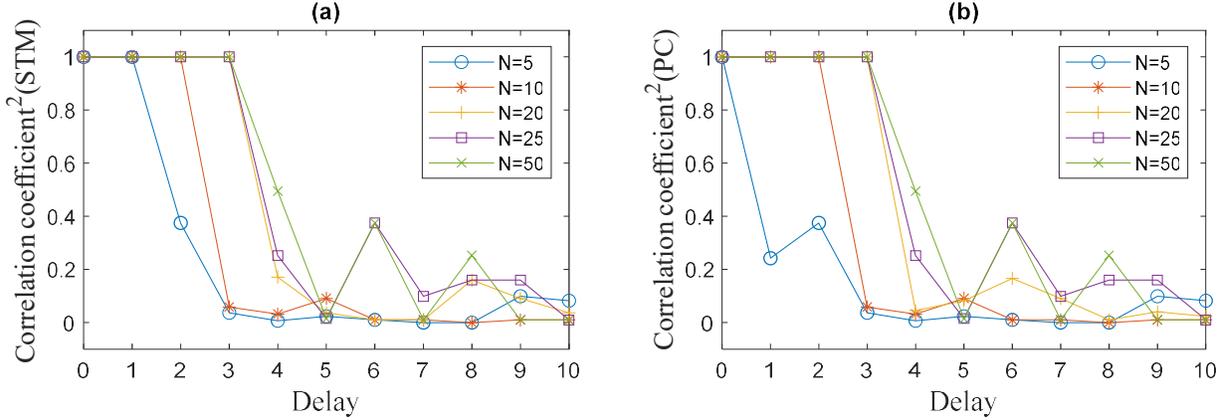

Fig. 3: (a) Square of correlation coefficient, $r^2_{STM}(d)$ for STM task in terms of delay $(d)$ and number of virtual nodes, (b) square of correlation coefficient, $r^2_{PC}(d)$ for PC task in terms of delay $(d)$ and number of virtual nodes.

Fig. 3(a) shows the square of the correlation coefficient, $r^2_{STM}(d)$ between the training data of STM task, $y^{n,d}_{STM}$ and output data, $y_{out}$ as a function of delay from $d=0$ to $d=10$. Each of the time steps correspond to a 10 ns delay. The STM correlation coefficient$^2$, $r^2_{STM}=1$ for all the number of virtual nodes, $N$ in consideration at delay, $d=1$ and starts to decrease with the increase of the delay. The $r^2_{STM}(d)$ tends to be higher in general with the increase in the number of the virtual nodes. Similarly, Fig. 3(b) presents the square of the correlation coefficient, $r^2_{PC}(d)$ between the training data of PC task, $y^{n,d}_{PC}$ and output data, $y_{out}$ as a function of delay from $d=0$ to $d=10$. Similar trends as STM have been observed for the PC task, for the correlation coefficient, as a function of the number of virtual nodes, and delay.

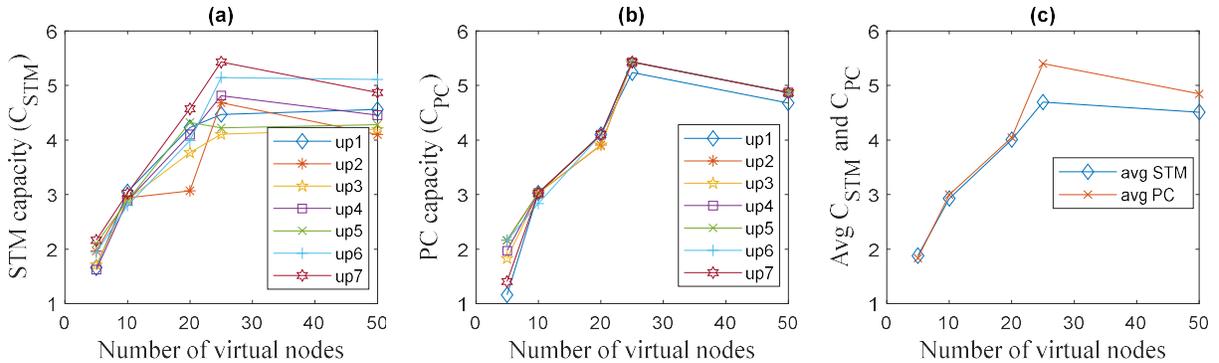

Fig. 4: (a) Short-Term Memory capacity $(C_{STM})$ and (b) Parity Check capacity $(C_{PC})$ of the reservoir as a function of the number of virtual nodes (c) Average STM and PC capacity of the reservoir as a function of the number of virtual nodes.

The dependency of STM capacity $(C_{STM})$ on the number of virtual nodes, $N$ in each signal is shown in Fig. 4(a). There is a general tendency of increasing STM capacity with an increasing number of virtual nodes for all the output nanomagnets. Fig. 4(b) shows the PC capacity ($C_{PC}$) vs. virtual node numbers (N) follows similar characteristics as STM task. The maximum capacity achieved by the reservoir for both STM and PC tasks is 5.43 bits for the case with output nanomagnet op7 and 25 virtual nodes. The obtained STM and PC capacities are comparable or higher than the other spintronic reservoirs[10,23,26,34]. The average STM and PC capacity of seven output nanomagnets are shown in Fig. 4(c) in terms of the virtual node numbers. The reservoir has an average STM capacity of ~4.69 and PC capacity of ~5.39 bits.

To separate the role played by the nonlinear nanomagnet reservoir in achieving the high STM and PC over that due to pre-processing (carrier amplitude modulation) and post-processing (filtering the carrier) we perform the following study. The pre-processed input is fed into a single-layer perceptron (SLP) network and its output post-processed before classifying and this is compared to the case of the reservoir with pre and post-processing. The result shows a correlation ($r^2$) of 1 for both STM and PC tasks, at delay 1 but very low or almost no correlation ($r^2$) for delay 2 and higher compared to the case with filters and reservoir as shown in Fig. 5. The calculated STM and PC capacities of the SLP are ~1.44 bits and ~1.43 bits, respectively while STM and PC capacities of the reservoir are ~3.52 bits and ~3.46 bits, which indicates the effectiveness of the reservoir over merely pre and post- processing.

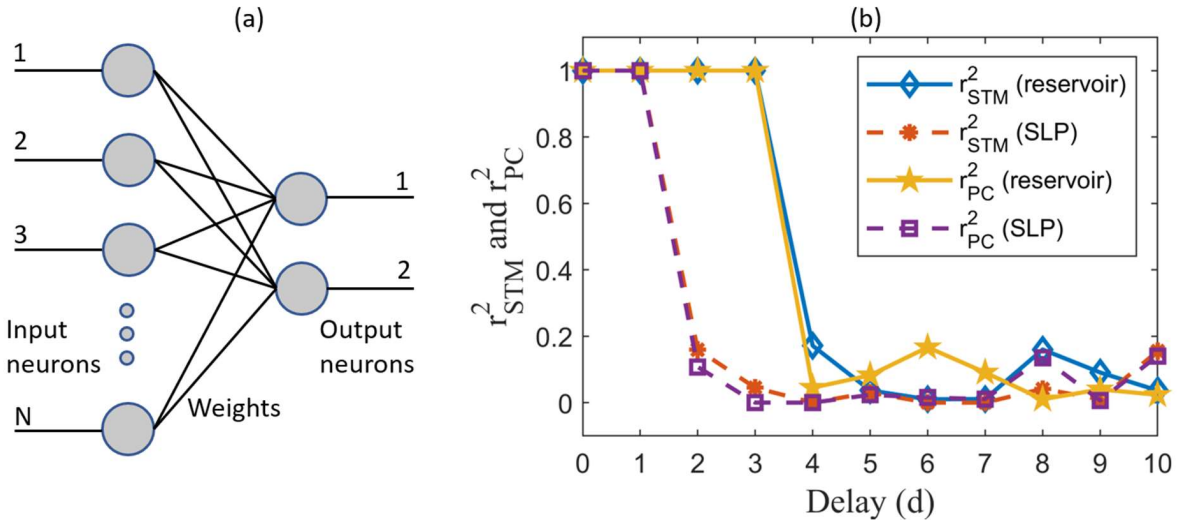

Fig. 5: (a) Architecture of single layer perceptron (SLP) which is used to measure the influence of pre and post-processing on the performance of reservoir (b) Calculated coefficient, $r^2_{STM/PC}(d)$ for STM and PC tasks with various delay, *d* for two different cases: with reservoir (solid lines) and without reservoir (only SLP and dashed lines). The number of virtual nodes for both cases are, *N*=20. The lines are drawn to guide the eye.

The total energy dissipation in the proposed reservoir system solely depends on the SAW excitation[35-37] as there is no other input mechanism needed. To estimate the energy consumption of the nanomagnets we also assume total generation of SAW induced strain from the piezoelectric substrate to the nanomagnets. The energy dissipated by the focused SAW IDT per input time period is $\left(\frac{P}{W}\right)Wt_{saw} = 0.87\ pJ$. The MTJ read out typically costs much less than 1 fJ[43-44] per virtual node and the passive filter costs no additional energy. Therefore, the readout cost of 50 virtual nodes is less than 50 fJ (0.05 *pJ*), meaking total RC energy cost 0.92 *pJ* per input time period. The spintronic reservoir energy consumption is compared with an equivalent CMOS-based echo-state network (ESN). The ESN is simulated to obtain the similar performance of nanomagnet RC. To achieve similar parity check capacity the CMOS ESN needs 11 neurons, which corresponds to 338 *pJ* of energy that is two-orders higher than the proposed nanomagnet reservoir (<1 *pJ*). The detailed energy dissipation and comparison is presented in the supplementary. Although CMOS ESN is able to achieve comparable accuracies for PC task, the STM task accuracies of the CMOS reservoir are still significantly low compared to our spintronic reservoir, which exhibits high capacities for both STM and PC tasks. The energy dissipation can be further decreased by applying higher frequency (> 4GHz) focused SAW, reducing the period of sine/square wave, and carefully selecting or optimizing material parameters. Furthermore, this NMRC scheme requires readout of only a single MTJ and enables the implementation of less external circuitry with more energy saving.

In summary, we have introduced a spintronic physical reservoir where a focused SAW is applied to the input. The non-linear response of the output nanomagnets are processed and output weights are trained through simple linear regression. The reservoir is able to identify sine and square waves with 100 % accuracy. In addition, we have demonstrated the expressivity of the reservoir by evaluating two figures of merit for RC. We have achieved average capacities of ~4.69 and ~5.39 for STM and PC respectively, which are indicative of a viable physical reservoir. The reservoir is extremely energy efficient and potentially needs two-orders of magnitude less energy than a CMOS-based ESN. Finally, the ability to use high-frequency SAW makes the device scalable to small dimensions, while the ability to modulate the envelope at a lower frequency (100 MHz) adds flexibility to encode different signals beyond the work in this paper. This could be key to applications such as speech recognition, anomaly detection, etc. using in-situ learning in edge devices.

## ACKNOWLEDGEMENT

M.F.F.C., W.A.M., M.M.R., and J.A. are supported in part by the National Science Foundation grant CCF-1815033 and Commonwealth Cybersecurity Initiative (CCI).

## AUTHOR DECLARATIONS

The authors have no conflicts to disclose

## SUPPLEMENTARY MATERIAL

See supplementary material for details of the reservoir computing method, energy dissipation of SAW, and energy dissipation of CMOS-based echo-state network.

## DATA AVAILABILITY

The data that support the findings of this study are available from the corresponding author upon reasonable request.

# Supplementary Material

# Focused Surface Acoustic Wave induced nano-oscillator based reservoir computing

Md. Fahim F. Chowdhury[1], Walid Al Misba[1], Md Mahadi Rajib[1], Alexander J. Edwards[2], Dhritiman Bhattacharya[3], Joseph S. Friedman[2], Jayasimha Atulasimha[1*]

[1]Department of Mechanical and Nuclear Engineering, Virginia Commonwealth University, Richmond, VA
[1]{chowdhurymf, misbawa, rajibmm, *jatulasimha}@vcu.edu
[2]Department of Electrical and Computer Engineering, The University of Texas at Dallas, Richardson, TX
[2]{alexander.edwards, joseph.friedman}@utdallas.edu
[3]Department of Physics, Georgetown University, Washington, DC
[3]dhritiman.bhattacharya@georgetown.edu

**Reservoir computing (RC):**

The reservoir computing is implemented as follows:

$$i_s(n) = \begin{cases} 0, & sine \\ 1, & square \end{cases} \quad (S1)$$

$$i_s = \{0,1,01, \ldots\ldots\ldots,1,1,0\}; \quad n \in \{1,2,3\ldots,n_{train},n_{test},\ldots\ldots,n_{max}\} \quad (S2)$$

$$s^n = [s_1^n\ s_2^n\ s_3^n\ \ldots\ldots\ s_N^n] \quad (S3)$$

Here $i_s$ is the input label of 100 ($n_{max}$) random sine or square waves. In each period of sine or square, the non-linear magnetization response (or magnetoresistance due to magnetization orientation of the soft layer of the MTJ) of the reservoir is read N times in an interval $\tau$, where $\tau = \frac{T}{N}$, $T$ = period of sine or square, and $N$ = number of virtual nodes. Here, $s^n$ is the virtual node vector of a sine or a square signal and the measured virtual nodes represent the states of the reservoir nanomagnets that are obtained from the output envelopes.

The current state of the reservoir depends on the current input and the previous state of the reservoir, which represents the short-term memory of the reservoir.

$$s^n(N+1) = f[s^n(N), i_s(n)] \quad (S4)$$

The optimum weights are obtained by linear Moore-Penrose pseudo-inverse operation to the training data. The optimized output weight is called learning and used to classify the test waveforms. The mean square error (MSE), the optimized weight matrix ($W_{out}$) is expressed as:

$$MSE = \frac{1}{n_{train}} \sum_{n=1}^{n_{train}} (y_{train}^n - W_{out}^T s^n)^2 \quad (S5)$$

$$W_{out} = y_{train}^t * pinv(s^t)\ ; \quad t = \{1,2,3\ldots,n_{train}\} \quad (S6)$$

Here $pinv$ finds the Moore-Penrose pseudoinverse of a matrix. Suppose the $W_{out}$ thus evaluated is:

$$W_{out} = [W_1, W_2, ..., W_N] \tag{S7}$$

Then the output of the reservoir (denoted as $y_{out}$) is obtained by the matrix multiplication of the learned weight and reservoir state or test data of the network.

$$y_{out} = W_{out}s^v \ ; \ v \in \{n_{test}, ..., n_{max}\} \tag{S8}$$

$$y_{out} = [W_1, W_2, ..., W_N] \begin{bmatrix} s_1^{n_{test}} & s_1^{n_{test}+1} & \cdots & s_1^{n_{max}} \\ s_2^{n_{test}} & s_2^{n_{test}+1} & \cdots & s_2^{n_{max}} \\ \vdots & \vdots & & \vdots \\ s_N^{n_{test}} & s_N^{n_{test}+1} & \cdots & s_N^{n_{max}} \end{bmatrix} \tag{S9}$$

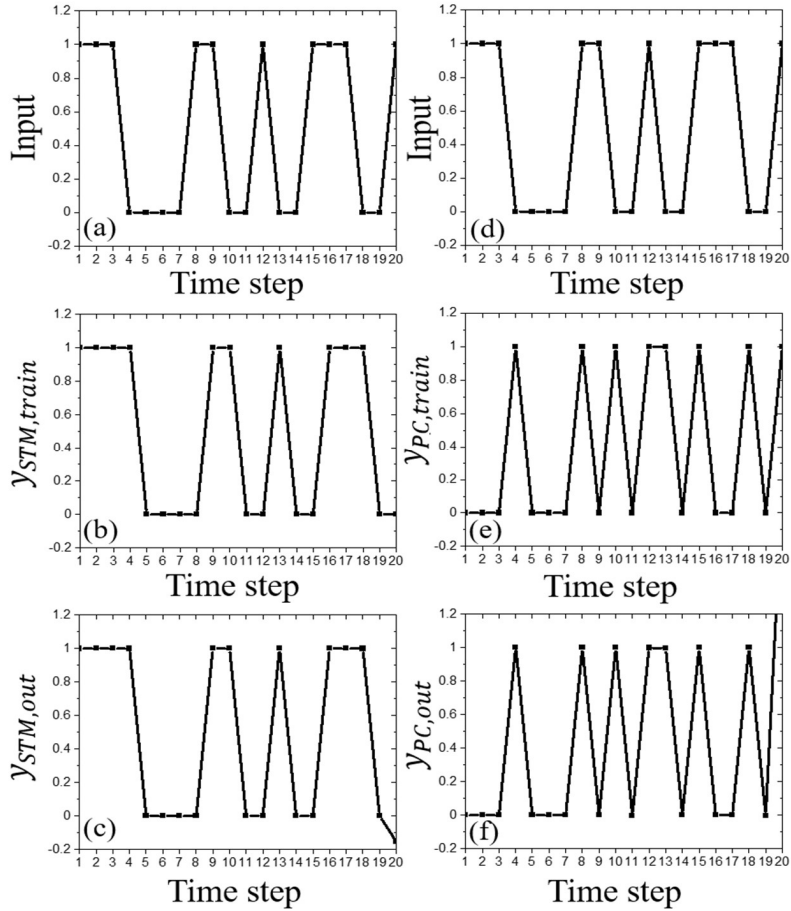

Fig. S1: The SAW input data ($i_s$) for STM (a) and PC (d) tasks. The training sequence of the input for (b) short-term memory task ($y_{STM,train}$) at delay, $d_{STM} = 1$ and (e) parity check task ($y_{PC,train}$) at $d_{PC} = 1$. The (c) STM output data ($y_{STM,out}$) and PC output data ($y_{PC,out}$). The virtual node number, $N = 20$. The output nanomagnet $op_3$. NOTE: The $op_3$ is merely used as an example, $op_7$ provided better STM and PC.

**Training and testing for short-term memory (STM) and parity check (PC) tasks:**

The input data ($i_s$), training data for STM ($y_{STM,train}$), and PC ($y_{PC,train}$) tasks, and the corresponding output data ($y_{STM,out}$ and $y_{PC,out}$) are shown in Fig. S1. The training data for the STM (Fig. S1(b)) and PC tasks (Fig. S1(e)) are defined in Equations (4) and (5) of the main article, respectively. The output is calculated using the magnetization dynamics of the output nanomagnet, $op_3$ and the number of virtual nodes per signal, $N = 20$. The optimized weights, $W_{out}$ are obtained using Equation (S6). The output data fits the training data for both STM and PC tasks at the delay, $d_{STM} = 1$ and $d_{PC} = 1$, respectively, which corresponds to the square of the correlation coefficient, $r^2 = 1.0$.

**Energy consumption of the SAW induced reservoir:**

To introduce strain in the magnetostrictive free layer of the input nanomagnet, we have applied 4 GHz SAW on the piezoelectric substrate through focused IDT (FIDT)[S1] with circular-arc focal points. The FIDT generates concentrated SAW energy with high intensity, which is localized on the center of the IDT, and produces higher amplitude waves[S2]. The energy dissipation due to SAW excitation on the piezoelectric substrate mostly depends on the potential ($V_s$) applied to induce required stress ($\sigma$) to strain the magnetostrictive free layer ($CoFe$) of the input MTJ, Young's modulus of CoFe ($Y$) the FIDT beamwidth ($W$), frequency of SAW ($f_{SAW}$), and the piezoelectric substrate ($LiNbO_3$) properties such as $d_{33}$ coefficient (ratio of induced strain to the applied electric field), admittance ($y_a$), SAW propagation speed etc. The maximum required stress is: $\sigma = \frac{\Delta stress\ anistropy}{\frac{3}{2}\lambda_s} = 186$ MPa, where saturation magnetostriction of CoFe, $\lambda_s$=250 ppm and the maximum change of magnetic anisotropy is $7.0 \times 10^4$ Jm$^{-3}$. The power dissipation by the SAW FIDT is defined by[S3]

$$\frac{P}{W} = \frac{1}{2}|V|^2 \left(\frac{y_a}{\lambda}\right) \tag{S10}$$

The required surface potential is determined by $V = \frac{\sigma D}{2Yd_{33}\sin\frac{\pi D}{\lambda}} = 3.87\ V$, where $Y = 200\ GPa$, $\sigma = 186\ MPa$, $d_{33} = 34.45\ pm/V$. For wavelength, $\lambda =897$ nm and admittance, $y_a = 0.21 \times 10^{-3}$ (S), the power dissipation, $\frac{P}{W} = 1757.58$ W/m.

If the IDT beamwidth ($W$) is 50 nm and SAW application time ($t_{SAW}$) is 10 ns, the energy dissipated per input is $\left(\frac{P}{W}\right)Wt_{saw} = 0.87 \times 10^{-12}$ J.

Table SI. Parameters related to energy calculation.

| SAW frequency, $f$ | 4 GHz |
|---|---|
| Piezoelectric constant, $d_{33}$ | 34.45 pm/V |
| Youngs modulus, $Y$ | 200 GPa |
| Admittance, $y_a$ | 0.21x10$^{-3}$ S |
| Required voltage, $V$ | 3.875 V |
| FIDT width, $W$ | 50 nm |
| SAW application duration, $t_{SAW}$ | 10 ns |

**Energy of CMOS-based echo-state-network:**

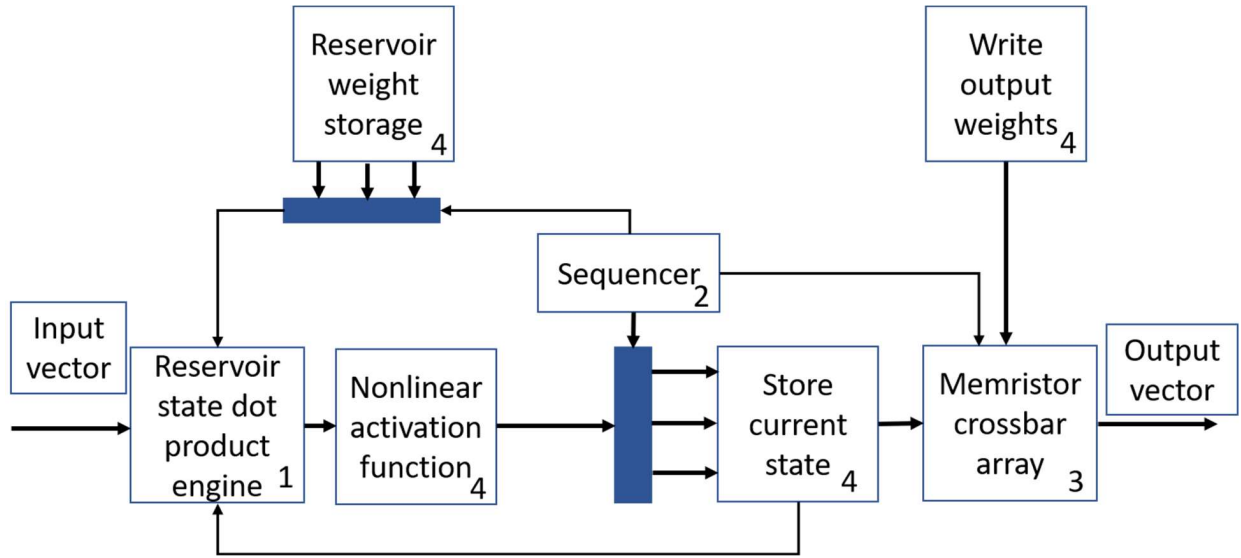

Fig. S2: The block diagram of digital reservoir system (CMOS-based echo-state network).

The reservoir energy dissipation is compared with an equivalent CMOS-based echo-state network (ESN) shown in Fig. S2. The arrows represent data path of the digital reservoir. The labeling of the blocks is as follows: 1 – arithmetic block, 2 – control logic block, 3 – memristor crossbar array, and 4 – memory block. Primary inputs are concatenated with the previous reservoir state and multiplied by a 16-bit fixed point reservoir weight matrix. This multiplication is performed row-by-row and then a non-linear activation function (Look Up Table) is used to calculate the reservoirs' internal state. Finally, in order to compare resource consumption of only the reservoir implementations and not the output layers, calculations for AEDP assume a memristor crossbar array is used as the output layer. The energy usage breakdown of the CMOS RC is given in Table SII.

Table SII. Energy usage breakdown of CMOS-based ESN

| Total CMOS energy | 169 pJ |
|---|---|
| Look up table energy | 81.1 pJ |
| Arithmetic units' energy | 87.9 pJ |
| Memristor Crossbar Array (MCA) energy | 0.0556 pJ |

**References:**

[S1] M. S. Kharusi and G. W. Farnell, Proceedings of the IEEE 60, 945–956 (1972).

[S2] M. B. Mazalan, A. M. Noor, Y. Wahab, S. Yahud, and W. S. W. K. Zaman, Micromachines 13, 30 (2021).

[S3] S. Datta, (Prentice Hall, 1986).